\documentclass[12pt]{article}
\usepackage{amssymb,amsmath}
\usepackage{array}
\usepackage{float}
\usepackage{graphicx}
\usepackage{subcaption}
 
\def\'#1{{\accent19\ifx #1i \i\else #1\fi}}

\def\be{\begin{equation}}
\def\ee{\end{equation}}
\def\bea{\begin{eqnarray}}
\def\eea{\end{eqnarray}}

\newcommand{\boldmathgamma}{\mbox{\boldmath$\gamma$\unboldmath}}

\newcommand{\boldmathnabla}{\mbox{\boldmath$\nabla$\unboldmath}}

\newbox\Ancha
\catcode`@=11
\newdimen\ex@
\title{ Compositeness,  Bargmann-Wigner  solutions within a U(1)-interaction quantum-field-theory expansion, and charge    }
 \author{J. Besprosvany$^*$ }
\date{Instituto de F\'{\i}sica, Universidad Nacional Aut\'onoma de M\'exico,
Apartado Postal 20-364,  01000, Ciudad de M\'exico, M\'exico }

\begin{document}


\maketitle

\jot = 1.5ex
\def\baselinestretch{1.9}
\parskip 5pt plus 1pt


\begin{abstract}

New solutions of  the Bargmann-Wigner equations     are obtained:  free fermion-antifermion pairs, each satisfying Dirac's equation,  with parallel momenta and momenta on a plane, produce vectors satisfying  Proca's equations.
     These equations are consistent with  Dirac's and  Maxwell's equations, as zero-order conditions within  a Lagrangian expansion for the  U(1)-symmetry quantum field theory.
  Such vector solutions' demand that they satisfy Maxwell's equations  and
 quantization fix the  charge. The current  equates   the vector field,  reproducing  the superconductivity    London equations, thus, binding
 and screening conditions.  The derived vertex  connects to QCD superconductivity and
  constrains four-fermion interaction composite models.











\end{abstract}

$^*$ email: bespro@fisica.unam.mx

\baselineskip 21pt\vfil\eject \noindent

\section{Introduction}

The Standard Model\cite{Glashow}-\cite{Salam} (SM)   accurately describes elementary particles and their hypercharge ($Y$),  weak (left-handed, $L$) and color ($c$) interactions, defined by the gauge groups  U(1)$_Y\times$SU(2)$_L\times$SU(3)$_c$, respectively; yet, puzzles remain as the origin of phenomenological constants, like the interactions' coupling constants.

Insight on these was provided on  proposed structures that generalize physical features: grand-unified theories\cite{Georgi}  assume a common group for the interactions, requiring a unique coupling constant at the unification scale, and setting constraints on their values.

 Compositeness, which  refers to related   properties   in systems  built from simpler ones,  underlies many physical systems and   provides information on them.
 The quark model\cite{Neeman} is  one  of its paradigms, as hadron features are derived from the quarks'. Likewise,  theories with more fundamental elements were proposed to explain SM features\cite{Harari,Fritzsch}; supersymmetry generalizes the SM composite quantum numbers   to additional feasible fermions and bosons\cite{Wess}; and  SM  structures can predict such constants\cite{BesproCoupling,neoPRD}, all suggesting    elementary composite configurations may provide  clues, for which we review relevant physical and formal setups.







 In superconductivity,
compositeness is also present.
  For its   Bardeen-Cooper-Schrieffer's\cite{Schrieffer} (BCS) theory, the relevant degree of freedom is a Cooper pair conformed of an electron and a hole. In addition, interaction screening allows   for free-particle behavior, which may enable  elementary-particle    properties to become manifest.


An   early application of this scheme in particle physics\cite{NambuJona}    describes a  mass-generating mechanism  for fermions and composite bosons, relating their masses, through a hypothesised interaction  built   from  four fermions, with binding and self-consistency features. A correspondence was argued between  such  a composite model and the U(1) theory\cite{Bjorken}. The  idea of  a quark condensate generating the Higgs\cite{Bardeen, MIRANSKYMay}, with a binding effect analogous to superconductivity, remains feasible.
Such a mechanism is dynamical, as  for radiative corrections\cite{EWeinberg}.

 We use   the simpler          Abelian   U(1)-group gauge symmetry, relevant at all ranges.  With focus
on fundaments,  a  gauge-interaction quantum field theory is a useful framework as a  four-vector's spin-1 degree of freedom can be constructed in terms of two  fermions'  spin-1/2.   The  U(1) interaction  picks a particular four-fermion interaction as effective theory for the Nambu-Jona Lasinio model\cite{NambuJonaII};  it comprises physically the hypercharge and  the electromagnetic interactions, and an effective version for the strong interaction that, e. g., models the quark anti-quark potential\cite{quarkantiquark}.

  Framed within the U(1) theory,   the  Bargmann-Wigner equations\cite{BargmannWigner} depart from  a fermion-antifermion pair, each element satisfying Dirac's equation,
that conforms a current element satisfying Proca's,  all comprising a
  composite   configuration.

In this paper,    we present a new class of   Bargmann-Wigner solutions, which signifies  their  space is dimensionally larger.    We  show that these equations   constitute zero-order terms, within a generated
 expansion for the Lagrangian, for a U(1)-interaction quantum field theory, constituting a compositeness limit.   For such   systems, we  show that Maxwell's equations  and    quantization    fix  the coupling constant.
 Such configurations satisfy superconductivity conditions through the London equations, connecting the current and the U(1) vector. Finally, we argue that this setup applies to QCD superconductivity, among other systems.
 The material is organized thus: Section II develops a Lagrangian expansion that contains the free Dirac's and Maxwell's equations as zero-order conditions;
   in  Section III, these are connected to the Bargmann-Wigner equations,  presenting more solutions.  Section IV derives a charge from Maxwell's equations consistency and quantization. Section V shows that the  London's equations are implied, linking
 fermion-antifermion states to superconductivity. Section VI analyzes other applications and draws conclusions.  Appendixes A-F present calculations in detail; notably, Appendix B contains
   Bargmann-Wigner  solutions with fermion-antifermion same-momentum and arbitrary spin configurations, and with combinations of  
   momenta on the same plane, including a confined configuration.  Appendix C     presents the adapted Bargmann-Wigner equations used.

\section{U(1) expansion}







A  U(1)-symmetry imposes four-vector field  $A_ \mu(x)$  minimal substitution  on Dirac's equation\cite{Dirac},
 setting its interaction with the fermion  represented by $\psi $, a   four-complex column  wave function accounting for spin and the antiparticle, with the  U(1)  four-current  given by $g j_\mu= g\bar\psi  \gamma_\mu \psi$,
with coupling $g$. The $x$ space-time coordinate dependence is hence usually omitted.
The U(1) Lagrangian  ${\cal L}(A _\mu, \psi )= {\cal L}_D( \psi )+   {\cal L}_K(A _\mu)+ {\cal L}_I(A _\mu, j_\mu )$ contains the  free-Dirac ($D$), vector-kinetic ($K$) components and  interaction term ($I$)
    ${\cal L}_I(A _\mu, j_\mu ) =- g  A_ \mu j^\mu$.
Defining the the lowest-order   solution through an external c-number  current\cite{Itzykson}   $j^s _\mu$      leads to    \begin{equation}\label{LagrangianC}
{\cal L}_C(A _\mu, \psi )={\cal L}(A _\mu, \psi )-{\cal L}_I(A _\mu, j^\mu )+{\cal L}_I(A _\mu, j_\mu^s )={\cal L}_D( \psi )+   {\cal L}_K(A _\mu)+{\cal L}_I(A _\mu, j_\mu^s )\end{equation}  as zero-order term,  with corrections ${\cal L}_I(A _\mu, j_\mu)-{\cal L}_I(A _\mu, j_\mu^s )$.

Next, we obtain the equations of motion to this order, showing that extended  Bargmann-Wigner   equations connect  to them, allowing for a charge definition, and implying that  $A_{\mu} $   satisfies
the superconductivity\cite{Schrieffer}  London's equations.
 We shall use the relativistic-quantum-mechanics framework  and the quantization/normalization condition.
\section{Dirac's and Maxwell's equations connection to Bargmann-Wigner equations }


The Euler-Lagrange equations for ${\cal L}_C(A_\mu, \psi )$ in Eq. \ref{LagrangianC} imply Dirac's free-particle equation for a U(1)-charged fermion\footnote{For the speed of light and reduced Planck's constant,
 $c=1$, $\hbar=1$  is assumed, respectively,  unless  needed.}\begin{equation} \label{DiracEq}
 ( i\partial_{\mu}     \gamma^{\mu}-m ) \psi(x)=0,
\end{equation}
with  $m$   its mass, $\gamma_\mu$    4$\times$4 matrices satisfying the Clifford algebra
$\{\gamma_\mu,\gamma_\nu \}=2 I  g_{ \mu\nu}$, transforming under the Lorentz group as vectors, $I$  the  identity matrix, and $g_{ \mu \nu}$   the metric.\footnote{The metric is defined
${g}_{\mu\nu}  =(1,-1,-1,-1)$. } These matrices are given in the Dirac representation in     {Appendix A}, with solutions in {Appendix B}.

  Now, with the  current  notation  ${j^s}^\mu\rightarrow j^\mu$, not specifying the zero order,
variations of $\cal L_C$  over $A_\mu$ lead to Maxwell's equations
 \begin{equation} \label{Maxwell}
 {\partial}^\nu  ( {\partial}_\nu A^s_\mu      -{\partial} _\mu  A^s_\nu )
  =  4 \pi  g     j_{\mu} ,
\end{equation}     where $A^s_\mu$  provides
       the solution to Maxwell's equations at this order.
The  homogenous equations are also implied  for $F_{\mu\nu} =       {\partial} _\mu    A^s_{\nu}-{\partial}_\nu {  A^s}_{\mu}$
\begin{equation}\label{MaxwellHomo}
  \frac{1}{2}\partial ^\mu\epsilon_{\mu  \nu \eta \sigma}{F}^ {\eta\sigma}=0.
\end{equation}

 The Bargmann-Wigner\cite{BargmannWigner}  equations  (applied for   two quanta,    {Appendix C}) comprise  two identical  spinors satisfying each Dirac's equation \ref{DiracEq}. These  conform a  vector satisfying   in turn
Proca's equations\cite{Proca} (which generalize Maxwell's  equations to the massive case)
\begin{equation} \label{semiMaxwell}
(\hbar c)^2 [{\partial}^\nu   ( {\partial}_\mu j_\nu  -{\partial}_\nu  j_{\mu} )]
  =(2 m c^2)^2 { j}_{\mu}.
\end{equation}

 The   current is thus equated to its derivatives
with   an explicit    Maxwell structure,\footnote{For arbitrary
$c$, $\hbar$ units, with $\partial_\mu=
(\frac{1}{c}\partial_0,  \boldmathnabla)$, $  j_\mu= (\bar{ \psi}  \gamma_0 { \psi}
,-c\bar{  \psi}  \boldmathgamma  { \psi} )$, $\boldmathgamma=(\gamma^1,\gamma^2,\gamma^3)$.}
and mass $2 m $, as $\partial^\mu j_\mu=0$ (Lorentz gauge).  These  equations translate   such spin-1/2 elements to  non-diagonal current components, for an electron and a positron\footnote{Within context, we use
the electron and its antiparticle as fermion representatives.} wave functions
$j_{\mu  }=\langle  \psi_c|\gamma_{0  }\gamma_{\mu  }|\psi\rangle$,    where $\psi_c=C \gamma_0  \psi ^*$,  $C=i\gamma^2\gamma^0$ is the  charge-conjugation matrix
 in the Dirac representation, for relevant cases ({Appendix C}). Fig. 1 represents these solutions, reinterpreted as a fermion-antifermion pair, as they connect to states in a Feynman diagram in which it converts to a massive vector.
 Given  the antisymmetric form
 $F^ j_{\mu\nu}={\partial}_\mu j_\nu  -{\partial}_\nu  j_{\mu}  $, Maxwell's homogeneous equations are satisfied, as in  Eq. \ref{MaxwellHomo}.
Same-momentum opposite spin, and   arbitrary energy, spin, and momentum on a plane,  extend the basis for larger kinematical regions, implying a feasible description of more than one fermion-antifermion pair (Appendix C).   For the fermion self-energy, the U(1) theory  leads to an effective (four-fermion) Hartree interaction,    which connects to the contact vector-vector term in ${\cal L}_I(A _\mu, j_\mu^s )$,
constituting an interaction case  in the Nambu-Jona Lasinio model\cite{NambuJonaII}.

%
\begin{figure}
\begin{centering}
\includegraphics[scale=0.7]{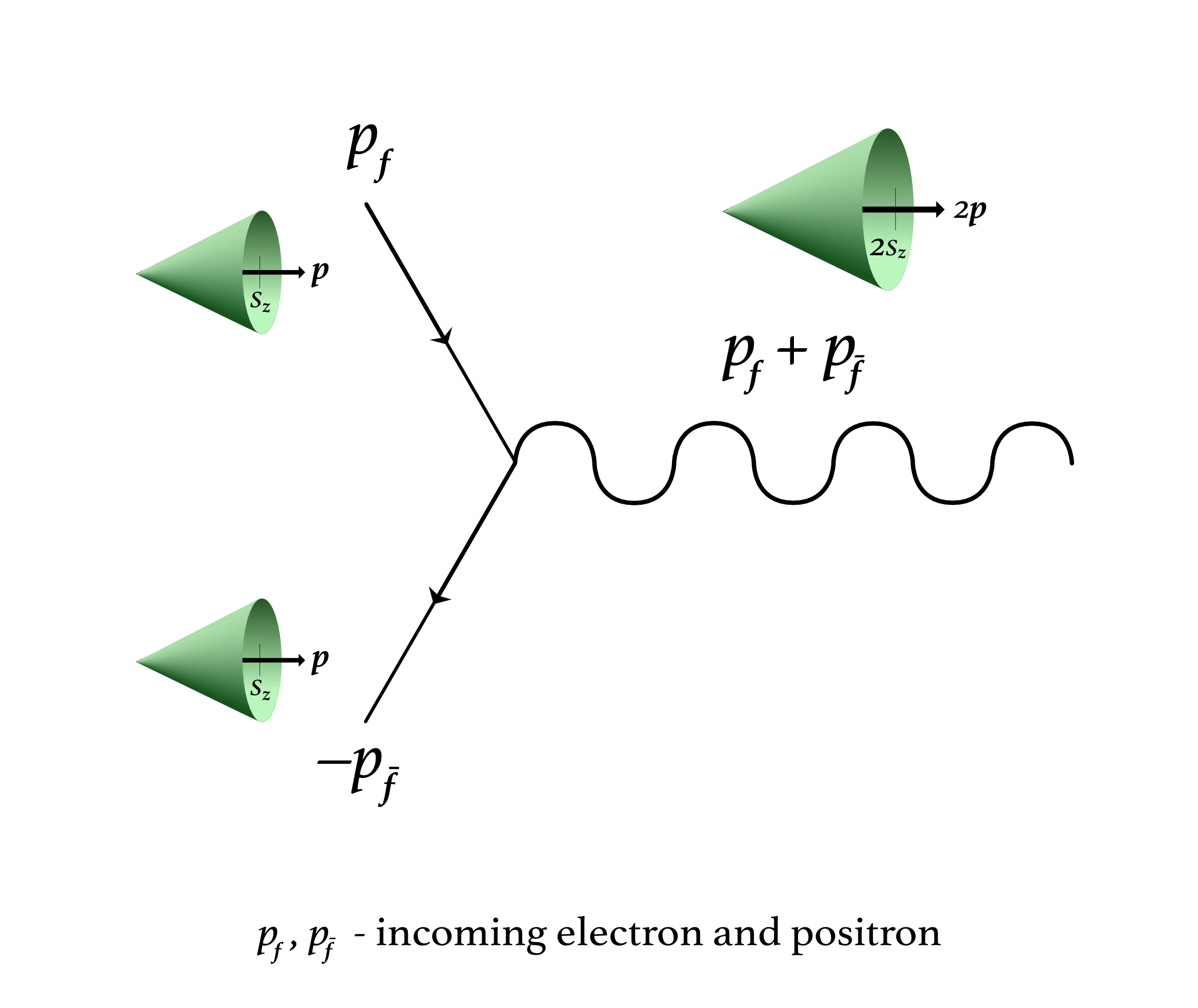}
\par\end{centering}
\caption{   Feynman   diagram  with vertex $ig\gamma_\mu$, related to solutions at fixed volume $V$,   as described in Eq. \ref{Maxwell}.
      Incoming     fermion-antifermion  $f  $-$\bar f  $   have  four-momenta         $  p _{ f}$, $p _{\bar f}$. The  Bargmann-Wigner solutions here have parallel momenta with   $  p _{ f}=p _{\bar f}=(E,0,0, p _z) $,    energy  $E = \sqrt{ m^2 + |{\bf p}|^2 }$,    ${\bf p}=(0,0, p _z)$,   each with spin $s_z$, represented by the small cones,
    producing     massive vector $A^s_\mu$ with  four-momentum $  p _{ f}+ p _{\bar f}$, polarization $2 s_z$.
   The inverse process represents vector-particle annihilation into two fermions.   }

%
 \label{FeynmannDiag}
\end{figure}

  For  the   field $A^s_{\mu} $ in Maxwell's equations \ref{Maxwell},  generated by  $ j_{\mu}$,
   the equations in \ref{semiMaxwell} are a necessary  consistency condition       restricting  $g$.
With their common  element $ j_{\mu}$ on the right-hand side, matching these equations
   implies   the former   is   the  latter times $4 \pi g/(2 m c^2 )^2$.
 We    rewrite Eq. \ref{semiMaxwell} in terms of  \begin{equation} \label{ApsiJpsi}  A^\psi_\mu=\frac{(\hbar c)^2}{(2m c^2)^2}\bar\psi   \gamma_\mu \psi.\end{equation}
 Using the Lorentz-gauge condition
\begin{equation} \label{KleinGordon}
  {\square} A^\psi_\mu=- \bar\psi   \gamma_\mu \psi ,
\end{equation}
 where the   ${\square}= \partial ^\mu\partial _\mu$      is the d'Alambertian.
Comparing the same elements, with derivatives acting upon  them,
leads to
\begin{equation} \label{AsScales}  4 \pi g A^\psi _\mu=   -A^s_\mu . \end{equation} 
This  equation ascertains $A^\psi _\mu$ is the  charge-independent component.

\section{Vector quantization and the U(1) charge}



 Quantum field theory imposes  field quantization, which identifies the factor
  connecting   $A^\psi_\mu$    with $A^s_\mu$  in Eq. \ref{ApsiJpsi}, as they constitute the {\it same} quantum. We  require the  normalization condition on conserved quantities as a proxy to quantization rules for
vectors\cite{Gasiorowicz}.

    A   vector  component extended to be massive, is analyzed in  {Appendix D}.
    The Maxwell's  energy-quantization condition\cite{Gasiorowicz,ProcaEnergy}, for Eq. \ref{Maxwell}, for such a vector with momentum $2 {\bf p}$, and mass $2 m$,  is\footnote{In the parallel momentum and spin case, the energy-momentum $T^{\mu\nu}$ satisfies   $\partial^\mu T_{\mu\nu}=0.$}  \begin{eqnarray} \label{energynormalizationMax}
 & \int d^3x\frac{1}{4\pi} [ \frac{1}{2}(  |{\bf E}^s  |^2 +   |{\bf B}^s |^2) +  \frac{(2 m)^2}{2}   ( 2 {A^s}_0   {A^s}_0  - g_{00}    {A^s} _\mu   {A^s}^\mu )]
= 2 E, 
 \end{eqnarray}
where $E= \sqrt{(mc^2)^2+ |   c\bf p| ^2}$, and we define    the electric and magnetic components from  ${A^s}^\mu =({A^0}^s,{\bf A}^s)$
    \begin{eqnarray}\label{ElectricMagnetic}
     {\bf E }^s&=& -\partial ^0 {{\bf A}^s}-{  \boldmathnabla} {A^0}^s  \\ \nonumber  {\bf B}^s&=&{\boldmathnabla}\times  { \bf A}^s ,\end{eqnarray}
valid
for a  U(1)    complex
  field, in correspondence with the Klein-Gordon equation case.

   On the other hand,   a Proca positive-energy free field with the same four-momentum  quantum-mechanical normalization in Eq. \ref{KleinGordon}
requires\cite{Greiner}
    \begin{equation} \label{normalization}
-\int d^3x {A^\psi_\mu}^*  \frac{i}{2 c } \overleftrightarrow{\partial}^0 { A^\psi}^\mu=1.
 \end{equation}

This massive-vector quantization  condition and that of the U(1) field connection in  Eq. \ref{energynormalizationMax}  implies, comparing their normalization constants ({Appendixes B, D})
 \begin{equation} \label{eleDiracVecEquality}
 \sqrt{4 \pi\hbar c} A^\psi _\mu=   -A^s_\mu .
 \end{equation}
Combining this equation  and the Dirac and U(1) vector relation in Eq. \ref{AsScales}  fixes $g$:
\begin{equation} \label{JaimeConstant}
 g/\sqrt{\hbar c }=   1/\sqrt{4 \pi},
 \end{equation}
whose value\footnote{{Appendix E} shows that this value is consistent with other unit choices.} is $g_{eLH}/\sqrt{\hbar c }=\sqrt{4 \pi}g/\sqrt{\hbar c }=1$ in Lorentz-Heaviside units.

 The         conserved property of the constructed  energy  and probability  in Eqs. \ref {energynormalizationMax}, \ref {normalization}, respectively, ensure that $g$ is Lorentz and gauge invariant.
 The obtained value,  $\alpha_g^{-1}=\hbar c/g^2= 4 \pi\simeq 12.6 $, is  between the couplings\cite{PDG}: weak $g_w$, hypercharge $g'$ and strong $g_s$ (associated to its U(1) effective form\cite{CoulombStron}) at low energies, consistently with unification conditions. Links to  SM interactions are feasible,    through a  unification model. The unification coupling constrains SM ones,  as obtained in other theories and energy scales\cite{Georgi,bespro9m1}.
One connection is through  quantum chromodynamics,  for its asymptotically-free behaviour\cite{asymptoticfree} relates to  a unification energy scale, with $g$ interpreted as the effective U(1) component\cite{CoulombStron}  sets  $(4/3)g_s^2=g^2$, implying  $\alpha_{g_s}^{-1}=\frac{16 \pi}{3}\simeq  16.76$ . Such   values are consistent   with the unified coupling constant in some models\cite{Wilczek}, with emphasis on those with compositeness\cite{CoyleWagner}.
  A generic U(1) interaction with   stationary coupling  running up to the unification scale, within feasible models\cite{unifiedcouplingmodels}    accounts for such scales  at high-energy. Such a value is also consistent with compositeness theories under fixed-point conditions for the coupling\cite{MIRANSKYMay}.

\section{Fermion-antifermion self-superconductor}


The Maxwell's equations   also relate   the current to the   vector potential.
Using Eqs.  \ref{ApsiJpsi}, \ref{AsScales}, and \ref{JaimeConstant}, extended London's equations are obtained
\begin{equation} \label{Jaime}
  j_\mu=-\frac{    g    \left ( \frac{  2 mc   }{\hbar } \right )^3 A^s_\mu }{ 2  m c},
\end{equation}
as  $A^s_{\mu} $ reproduces   superconductivity  behavior.
Paired particles  with  $ 2 m$  mass associate  this superconductivity condition  to the density $n_j=  \left ( \frac{2 mc    }{ \hbar  } \right )^3, $ with cube unit size of reduced Compton wavelength\cite{ComptonLength}  $ \lambda_C =  \frac{ \hbar   }{ 2  mc }   $.
We compare them using            the  generalized Gordon  identity\cite{Gordon} ({Appendix F}) for the interactive case
 \begin{equation} \label{Current}
\bar \psi i\overleftrightarrow{\partial}_\mu\psi-2 g \bar \psi \psi   A_{\mu}+
\frac{i}{2}{\partial}^\nu  \bar \psi  [\gamma_\mu,\gamma_\nu] \psi
  =2 m\bar\psi  \gamma_\mu \psi .
\end{equation}
 Such identities contain the Bargmann-Wigner equations in the compositeness limit.
 With proper normalization and multiplying by $g$,    the U(1) four-current $g j_\mu= g\bar\psi  \gamma_\mu \psi$  has\cite{Schrieffer},  on the left-hand side,    a paramagnetic contribution  (first term), and a diamagnetic (second) term, where
  ${\bf A} (x)$ is the U(1) vector potential,  reproducing for the current  the structure  of London's superconductivity equations\cite{London} (in the London gauge $ {\boldmathnabla}\cdot{\bf A}=0$)
  \begin{equation} \label{London}
g{\bf j} (x)=-\frac {{g}^2 {\bf A} (x) n_c}{   mc  },
\end{equation}
where $n_c$ is the electron density, to be compared with the local scalar bilinear $\bar \psi\psi$. Eq. \ref{London} is a valid  approximation  to Eq. \ref{Current} with only the second term on the left-hand side, in the regime of small wave-function variation, e. g.,
$|\bar \psi i\overleftrightarrow{\partial}_\mu\psi/\bar \psi \psi | \ll    |2 g A_\mu| $.
  For this  second superconducting component in Eq. \ref{Current},  the  U(1) field is    associated  to  the  system's length $ L $.
  Over long distances  $L \gg \lambda_C $, such component's  contribution
 is negligible     as compared to the third component, which contains the Maxwell's equations contribution in Eq. \ref{semiMaxwell}, ({Appendix F}) and   depends on  $\lambda_C$,  implying this   constitutes a sizable component in   a fermion's generated interactive field. As for the Meissner effect, a current is generated without friction, as a classical expansion of this equation manifests it. Unlike the usual London equations, which describe current under external fields, here they are self-consistent; also,   the current oscillates, not decays. Substitution of Eq. \ref{Current} into ${\cal L}_I(A _\mu, j_\mu^s )$ implies such a term is atractive, as the on-shell  $A _\mu$ ({Appendix D}) is space-like.


   This scheme applies to models and scenarios in which compositeness is assumed explicitly,  starting with the application of the BCS model within quantum field theory through the Nambu Jona-Lasinio model\cite{NambuJona};   SM  extensions   reproduce spontaneous symmetric breaking\cite{Bardeen, MIRANSKYMay} through quark condensates, with supersymmetry\cite{Wagner}, technicolor\cite{technicolor},   and some unification models.

\section{Applications and conclusions}
     Methods are added  to known cases as   extensions  \cite{bespro9m1,technicolor} requiring composite elements.
   Compositeness can be applied to  SM couplings\cite{BesproCoupling}, as  connections were   recently exploited to obtain information on the quark masses\cite{neoPRD}.
As the current is a particular   bilinear-spinor term, similarly  to     superconductor Cooper pairs\cite{Cooper}, a composite regime is feasible, under some conditions.  Generically, a 4-fermion interaction is   self-consistent,  as described in the  Hartree approximation\cite{Hartree}   or  as order parameter    in the   Landau-Ginzburg theory\cite{LandauGinzburg}.
          The generation of a
 boson quantum within superconducty\cite{Schrieffer, NambuJona}  suggests screening and binding conditions.

The consistency of the implied Bargmann-Wigner condition in Eq. \ref{semiMaxwell} with Maxwell's equations and quantization fixes the effective U(1) charge.  Derived  London's equations imply  superconductivity conditions, as a zero-order contribution  within an induced expansion.  Universal $m$-independent properties follow.

  The resulting fermion-antifermion
  pair with parallel momenta     induces
    a   $2 m$ vector resonance described by the  $A^s _\mu$  relation to the current through the extended London  equations
 \ref
 {Jaime}.   Such a vector   superconductive  state constitutes zitterbewegung\cite{zitter}, and  could be attained  by  stabilizing their interaction and with external fields\cite{Info}-\cite{opticaltrap}, recreating screening and binding conditions.  It should alter   ``fermonium" (fermion-antifermion)  properties, possibly  detectable in positronium (an e$^+$ e$^-$ system).

Such a vector resonance is  also consistent with the  interpretation of the expansion   through  a massive  contribution.
The   component ${\cal L}_m (A^{  \mu}) ={\cal L}_I(A^{  \mu} , A_\mu  )  =\frac{(2m)^2}{2} A_\mu A^\mu $ corresponds to
 a  $2 m$ mass term for ${A}^\mu$ in which one may    depart in an expansion that includes    a massive vector term:    ${\cal L}_D( \psi )+   {\cal L}_K(A _\mu)+ {\cal L}_m (A^{  \mu}) $ and  orders in  ${\cal L}_I(A _\mu, \psi )-{\cal L}_m (A^{  \mu}) $. Some models construct SM vectors fields dynamically and as composite objects\cite{Miransky}.
 The massive vector $A^s_ \mu$ (or $ j^\psi_ \mu$ ,) constructed  a free fermion-antifermion pair constitutes   an   $A _\mu$ component, within the zero-order Lagrangian expansion.










  Feasible connections arise for the U(1) compositeness regime, in limits studied in quantum electrodynamics (QED). In the infrared limit\cite{Appelquist}, reduced photon-interaction diagrams  create conditions for the prevalence of a   momentum-independent renormalized coupling term.
At high energy, QED predicts a fixed point\cite{Kogut}, for not too large couplings, in which  four-fermion interactions become renormalizable, allowing for
coupling momentum-independence, with applications as constructing realistic technicolor models\cite{technicolorGiudice}.
  In addition,   QCD's asymptotic freedom  make the paper's free-particle  solutions and conditions relevant for this regime.

 High density implies high-energy conditions, and under  asymptotic-freedom  for   QCD, free-particle behavior  and an attractive interaction generate superconducting conditions\cite{Alford}. These conditions are feasible in nuclear physics too\cite{Baym}. For Fermi liquids under QCD superfluidity or superconducting conditions,  at the
  Fermi-gas surface,   four-fermion vertices are relevant  with the kinematic structure associated with Cooper pairing (i.e. two quarks scattering with equal and opposite spatial momenta)\cite{Schafer, Schwetz}.
This paper's new solutions  provide  parametrizations for superconducting confined systems, sensible to boundary conditions\cite{Boundary}. In addition, its derived fermion-antifermion interaction components provides   matrix elements for Dirac-equation wave functions and their conjugates,
 relevant in the theory of superconductivity\cite{Bailin}.




 An expansion with a c-number current   constructed from a fermion-antifermion    is assumed, akin to a mean-field, with corrections to such  one-particle self-component solution extracted  from   the coupled Maxwell-Dirac equations,  producing the fermion-vector interaction. Solutions were found for the Dirac equation under a massive vector   free field\cite{VolkowMassive}, generalizing Volkow's solution for a massless vector.
Other corrections should consider second quantization,  many-particle  contributions,
  three-dimensional effects, loop corrections, and
  renormalization.


 Under compositeness conditions,   for a U(1) quantum field theory, the coupling is derived from   the  consistency
 of a quantized current from  fermion-antifermion free solutions, and Maxwell's equations.
 The  point of view emerges that such a constant  is inherent to a theory\cite{Einstein}. Superconducting conditions are induced, within
a generated expansion.
  The next task is to obtain  the  subsequent  corrections  to  such a constant.

{\bf Acknowledgements}

 The
author  acknowledges discussions with A. A. Santaella on the extended Gordon identities and Bargmann-Wigner equations applications, and  support from L. Novoa and M. Casa\~nas at the IF-UNAM with graphics, and from  the Direcci\'on General de Asuntos del Personal Acad\'emico,  UNAM,  through  Project IN117020.


\appendix
\renewcommand{\theequation}{A\arabic{equation}}
 \setcounter{equation}{0}

\subsection*{Appendix  A: Gamma-matrix identities and the Dirac representation}
 Among $4\times4$  $\gamma^\mu$       matrix properties, we list\cite{Itzykson}:

\begin{equation} \label{2gamiden}
 \gamma_\mu \gamma_\nu   = g_{\mu\nu} -i  \sigma_{\mu\nu}    \end{equation}
where $ \sigma_{\mu\nu}=\frac{i}{2}[\gamma_\mu ,\gamma_\nu ].$
\begin{equation} \label{3gamiden}
 \gamma_\mu \gamma_\nu\gamma_ \eta   = g_{\mu\nu}  \gamma_\eta   +g_{\nu\eta}  \gamma_\mu  -g_{\mu\eta}  \gamma_\nu
+i\epsilon_{\mu  \nu \eta \sigma}  \gamma_5 \gamma^\sigma,\end{equation}
   the chirality $\gamma_5=i \gamma^0\gamma^1\gamma^2\gamma^3$, and $\epsilon_{\mu  \nu \eta \sigma}$ is the four-dimensional Levi-Civita tensor:
 \begin{equation}\label{LeviCivita}\epsilon^{\mu  \nu \eta \sigma}=\left \{ \begin{matrix}
 1  &  {\rm if\ \{ \mu,  \nu, \eta, \sigma  \}\ is\  an\ even\  permutation\   of\  \{ 0,1, 2,3  \} } \\
 -1 &    \!\!\!\!\!\!\!\!\!\!\!\!\!\!\!\!\!\!\!\!\!\!\!\! \!\!\!\!\!\!\!\!\!\!\!\!\!\!\! \  \ {\rm if\  \{ \mu,  \nu, \eta, \sigma  \}\ is\ an\  odd\  permutation\ }\\
 0  &  \ \ \   \!\!\!\!\!\!\!\!\!\!\!\!\!\!\!\!\!\!\!\!\!\!\!\!\!\!\!\!\!\!\!\!\!\!\!\!\!\!\!\!\!
 \!\!\!\!\!\!\!\!\!\!\!\!\!\!\!\!\
  \!\!\!\!\!\!\!\!\!\!\!\!\!\!\!\!\!\!\!\!\!\!\!\!\!\!\!
 \!\!\!\!\!\!\!\!\!\!\!\!\!\!\!\!\!\!\!\!\!\!\!\!\! {\rm otherwise}
    \end{matrix} \right .
    \end{equation}

 The Dirac representation for the $\gamma^\mu$ matrices  is given next, where we use the $2\times2$ Pauli matrices,
\begin{equation}
\sigma^1= \left (\begin{matrix}
0 & 1 \\
 1  & 0
    \end{matrix} \right )
   \ \
   \sigma^2= \left (\begin{matrix}
0 &-i \\
 i  & 0
    \end{matrix} \right )
  \ \
    \sigma^3=\left ( \begin{matrix}
1 &0 \\
 0  & -1
   \end{matrix} \right )
  \end{equation}
and $I_{2\times2} $, the   identity matrix, to define them:
\begin {equation}
\gamma^0=\left (\begin{matrix}
   I_{2\times2} & 0 \\
0  & -I_{2\times2}   \end{matrix} \right ) \ \
\gamma^j =\left (\begin{matrix}
   0 & \sigma^j\\
-\sigma^j  & 0
\end{matrix} \right ) \ \  j=1,2,3.
\end{equation}

\subsection*{Appendix  B: Dirac's equation fermion-antifermion solution
combinations}

\renewcommand{\theequation}{B\arabic{equation}}
 \setcounter{equation}{0}

    Dirac's free equation  in \ref{DiracEq}
  contains particle       and antiparticle ($x$-dependent) solutions  $\psi_u(p,s_\alpha)$, $\psi_v(p,s_\alpha)$   with
    $p$,  $s_\alpha$,     the associated momentum, and   spin four-vectors, respectively.  Negative-energy solutions  are associated to antiparticles,
     with opposite  quantum numbers assigned within the  Dirac-sea interpretation, assumed in the notation, while second quantization provides  a symmetric and consistent treatment.  $p$ is chosen with spatial components along $\hat{\bf z}$:
   $p^\mu =(E,0,0,  p_z)=(\sqrt{p_z^2+m^2},0,0,  p_z)$.
 The wave functions are separated
   $\psi_u(p,s_\alpha)= u(p,s_\alpha) e^{-ipx}$,  $\psi_v(p,s_\alpha)= v(p,s_\alpha) e^{ ipx}$, where $u(p,s_\alpha)$,    $v(p,s_\alpha)$ are associated spin  states.
   We    construct   them from  \begin{eqnarray}\label{us} u_{1}= \left(\begin{array}{lcr}
  1\\  0\\ 0\\ 0
 \end{array}\right) , \ \ u_{2}= \left(\begin{array}{lcr}
  0\\ 1\\ 0\\ 0
 \end{array}\right) \\ v_{1}=\left(\begin{array}{lcr}
  0\\  0\\ 1\\ 0
 \end{array}\right), \ \ v_{2}=\left(\begin{array}{lcr}
  0\\  0\\ 0\\ 1
 \end{array}\right),
   \end{eqnarray}
 with \begin{eqnarray} \label{DiracSpinors}
    u(p,s_{+,-}) &=&\frac{1}{\sqrt{2   m(E+m)}  }( E \gamma_0 -p_z \gamma_3 +m)u_{1,2}  \\  \nonumber
     v(p,s_{+,-}) &=&\frac{1}{\sqrt{2   m (E+m)}  }(  -E \gamma_0+p_z \gamma_3 +m)v_{2,1} \\
  \end{eqnarray} classified by   the spin operator $-\frac{1}{2} \gamma_5 n\cdot\gamma\  p\cdot\gamma$, defined by the helicity vector
  $n=\frac{1}{m}(|{\bf p}|,E {\bf p}/|{\bf p}|)$:
  \begin{eqnarray} \label{SpinOpeAct}
   \frac{1}{2m}\gamma_5(p_z\gamma^0-E\gamma^3) u(p,s_+) &=&  \frac{1}{2} u(p,s_+)  \\  \nonumber \frac{1}{2m}\gamma_5(p_z\gamma^0+E\gamma^3) v(p,s_+) &=&  - \frac{1}{2} v(p,s_+), \\   \nonumber
    \nonumber \frac{1}{2m}\gamma_5(p_z\gamma^0+E\gamma^3) v(p,s_-) &=&   \frac{1}{2} v(p,s_-), \nonumber
  \end{eqnarray}
 where
  e. g. $u(p ,s_+)$ is a positive-energy $p_z$-momentum positive spinor  and     $v(p_z,s_-)$ is a   $p_z$-momentum negative spinor,   both along $\hat{\bf  z}$.
We use spinor
  normalization
   \begin{equation}\label{normalizationfer}\langle  u(p,s_\alpha) |  u(p,s_\alpha)\rangle
=  \langle  v(p,s_\beta) |  v(p,s_\beta)\rangle=E/m.
\end{equation}

 Relevant  current elements for    $u$- and $v$-type states,
 produce the   transverse and longitudinal polarization current matrix elements, respectively:
 \begin{eqnarray}\label{DiracCurrenttildefirst}
 j^{\mu++}_{   v u}   =    j^{\mu++*}_{u   v}  =    \langle \psi_v(  p,s_+)|\gamma_0 \gamma^\mu| \psi_u( p,s_+)\rangle
  & = & \frac{m}{E} e^{  - 2 i E t+ 2 i  p_z z } \left (0, 1  , i  ,0\right ) \label{DiracCurrenttildeTrans} \\ \label{DiracCurrenttildefirstend}
j^{\mu+-}_{    vu}   =   {  j^{\mu-+*}_{u   v}} = \langle \psi_v(    p,s_-)|\gamma_0 \gamma^\mu| \psi_u(p,s_+)\rangle
& = & \frac{-1   }{E} e^{- 2 i E t+ 2 i  p_z z}   \left (k_z,0,0,E  \right ) .\end{eqnarray}


{\bf  Bargmann-Wigner solutions on a plane }

A generalization is made for a stationary  fermion-antifermion  pair. We define   fermion states  with two  momenta on a  plane:       $p_{yz} =(E_{yz},0, p_y,  p_z)$, $E_{yz}=\sqrt{p_y^2+p_z^2+m^2}$  and        $ \tilde  p_{yz}  =(\tilde  E_{yz},0, \tilde  p_z, \tilde  p_y)$, $  {\tilde  E}_{yz}=\sqrt{{\tilde  p}_y^2+{\tilde  p}_z^2+m^2}$
and  possible  spin combinations:
The wave functions are
   $\psi_u(p_{yz},s_\alpha)= u(p_{yz},s_\alpha) e^{-i p_{yz}x}$,
    $\psi_v(p_{yz},s_\alpha)= v(p_{yz},s_\alpha) e^{ ip_{yz}x}$, where $u(p_{yz},s_\alpha)$,    $v(p_{yz},s_\alpha)$ are associated spin  states.
   We    construct them
 with \begin{eqnarray} \label{DiracSpinorsGen}
    u(p_{yz},s_{+,-}) =\frac{1}{\sqrt{2   m(E+m)}  }( E \gamma_0 -p_y \gamma_2 -p_z \gamma_3 +m) u_{1,2}   \\  \nonumber
      v(p_{yz},s_{+,-}) =\frac{1}{\sqrt{2   m (E+m)}  }(  -E \gamma_0+p_y \gamma_2+p_z \gamma_3 +m) v_{2,1} ,
  \end{eqnarray}
   with   $u_{1,2}$, $v_{1,2}$ in Eqs. \ref{us}. They are
    classified by the appropriate spin operator, along   $n_{yz}$
    $-\frac{1}{2} \gamma_5  n_{yz} \cdot\gamma p_{yz} \cdot\gamma$, defining, e. g.,
  \begin{eqnarray} \label{SpinOpeActyz}
  -\frac{1}{2} \gamma_5  n_{yz} \cdot\gamma p_{yz} \cdot\gamma u(p_{yz},s_+) =  \frac{1}{2} u(p_{yz},s_+) \\  \nonumber  -\frac{1}{2} \gamma_5  n_{yz} \cdot\gamma p_{yz} \cdot\gamma v(_{yz},s_-) =   \frac{1}{2} v(_{yz},s_-), \nonumber
  \end{eqnarray}

The fermion and anti-fermion wave functions
\begin{eqnarray}\label{feraniferwavefun}
   | M_u \ \rangle &=& a| \psi_u(  p_{yz},+)\rangle+b| \psi_u( \tilde p_{yz},+)\rangle+c| \psi_u(  p_{yz},-)\rangle+d| \psi_u( \tilde p_{yz},-)\rangle  \\ | M_v \ \rangle&=&  a^*| \psi_v(  p_{yz},+)\rangle- b^* | \psi_v( \tilde p_{yz},+)\rangle+ c^*| \psi_v(  p_{yz},-)\rangle-d^*| \psi_v( \tilde p_{yz},-)\rangle   \end
   {eqnarray}  produce  current matrix elements:
$ \langle M_v  | \gamma_0 \gamma^\mu| M_u \rangle= a^2 j^{\mu++}_{vu\ _{yz}}+b^2 j^{\mu++}_{{ \tilde  v  \tilde u\ _{yz}}}+c^2 j^{\mu--}_{vu\ _{yz}}+d^2 j^{\mu--}_{{ \tilde  v  \tilde u\ _{yz}}}
+ ac(j^{\mu-+}_{{ v   u\ _{yz} }}+  j^{\mu+-}_{{  u v\ _{yz} }})+ bd(j^{\mu-+}_{{ \tilde v \tilde u\ _{yz} }}+  j^{\mu+-}_{{\tilde v \tilde u\ _{yz} }})$,
 where $a$, $b$,  $c$, $d$ are arbitrary constants,
 producing  current matrix elements:
 \begin{eqnarray}\label{DiracCurrenttilde}    j^{\mu++}_{vu\ _{yz}} = \langle  \psi_v(  p _{yz},s_+)|\gamma_0 \gamma^\mu| \psi_u(  p _{yz},s_+)\rangle
  & = & \nonumber \\-\frac{e^{-2 i \left({E} t- p_y y- p_z z\right)}}{{E}}
 ( p_y,-i m,\frac{{E}^2+2 {E} m+m^2+p_y^2-p_z^2}{2
   ({E}+m)},\frac{p_y p_z}{{E}+m} ) \\
   j^{\mu++}_{{ \tilde  v  \tilde u\ _{yz}}} = \langle  \psi_v( \tilde p _{yz},s_+)|\gamma_0 \gamma^\mu| \psi_u( \tilde p _{yz},s_+)\rangle
     & = & \nonumber  \\ \frac{e^{-2 i \left({E'} t-  p'_y y-  p'_z z\right)}}{{E'}}
     ( p'_y,-im, \frac{{E'}^2+2 {E'} m+m^2+{p'_y}^2-{p'_z}^2}{2
   ({E'}+m)}, \frac{p'_y p'_z}{{E'}+m})   \\
   j^{\mu--}_{{   v  u\ _{yz}}}   =  \langle  \psi_v(   p _{yz},s_-)|\gamma_0 \gamma^\mu| \psi_u(   p _{yz},s_-)\rangle
    & = &\nonumber  \\-\frac{e^{-2 i \left({E} t- p_y y-  p_z z\right)}}{{E }}
    (p_y,im,\frac{{E}^2+2 {E} m+m^2+p_y^2-p_z^2}{2
   ({E}+m)},\frac{p_y p_z}{{E}+m}) \\
   j^{\mu--}_{{ \tilde  v  \tilde u\ _{yz}}}   =  \langle  \psi_v( \tilde p _{yz},s_-)|\gamma_0 \gamma^\mu| \psi_u( \tilde p _{yz},s_-)\rangle
    & = &\nonumber  \\ \frac{e^{-2 i \left({E'} t-  p'_y y-  p'_z z\right)}}{{E'}}
    ( p_y, im, \frac{{E'}^2+2 {E'} m+m^2+{p'_y}^2-{p'_z}^2}{2
   ({E'}+m)}, \frac{p'_y p'_z}{{E'}+m})   \\
  j^{\mu-+}_{{ v   u\ _{yz} }}   =   j^{\mu+-}_{{ v   u\ _{yz} }}   =    \langle \psi_v(   p _{yz},s_-)|\gamma_0 \gamma^\mu| \psi_u( p _{yz},s_+)\rangle
  & = & \nonumber  \\ -\frac{e^{-2 i \left({E} t-  p_y y-  p_z z\right)}}{{2 E}}
 (2 i p_z,0,\frac{2 i p_y p_z}{{E}+m},\frac{i \left({E}^2+2 {E} m+m^2-p_y^2+p_z^2\right)}{{E}+m})
     \\
j^{\mu-+}_{{ \tilde v \tilde u\ _{yz} }}   =  j^{\mu+-}_{{ \tilde v \tilde u\ _{yz} }} = \langle \psi_v( \tilde   p _{yz},s_-)|\gamma_0 \gamma^\mu| \psi_u(\tilde p _{yz},s_+)\rangle
& = & \nonumber\\  \frac{e^{-2 i \left({E'} t-  p'_y y-  p'_z z\right)}}{{2 E'}}
( 2 i p'_z,0, \frac{2 i p'_y p'_z}{{E'}+m}, \frac{i \left({E'}^2+2 {E'} m+m^2-{p'_y}^2+{p'_z}^2\right)}{{E'}+m} )        \label{DiracCurrenttildeend}       \end{eqnarray}
A U(1) vector is thus obtained from a fermion-antifermion pair. The  constants  $a$,  $b$,   $c$,  $d$ are set by the boundary conditions.

\subsection*{Appendix  C:
  Bargmann-Wigner equations}
  \renewcommand{\theequation}{C\arabic{equation}}
 \setcounter{equation}{0}

Eq. \ref{DiracEq}  is  written equivalently in the   transposed version
\begin{eqnarray} \label{DiracEqHerTrans}
 \psi^t  (   i\overleftarrow{\partial}_{\mu}  {\gamma^{\mu}}^ t-m )  =0   \\
  \psi^t C C^t (   i\overleftarrow{\partial}_{\mu}{\gamma^{\mu}}^ t-m ) C   =0 \\
  \psi^t C  (   i\overleftarrow{\partial}_{\mu}  {\gamma^{\mu}}+m ) =0 \label{DiracLeft}
\end{eqnarray}
In the Dirac matrix representation, the charge conjugation operator is $C= i\gamma^2\gamma^0$, which satisfies $C C^t=C^t C =-C^2= 1$,  $C^t {\gamma^\mu}^t C=-\gamma^\mu $.
The matrix pair configuration $\Psi_{\scriptscriptstyle{BW} }=\psi \psi^t C $ has only  symmetric elements\cite{Lurie},
  implying   $\Psi_{{\scriptscriptstyle{BW} }}= (j_{{\scriptscriptstyle{BW} }}^\mu \gamma_\mu+F_{{\scriptscriptstyle{BW} }}^{\mu\nu}   \sigma_{\mu\nu})C$, with $\sigma_{\mu\nu}$ defined after Eq. \ref{2gamiden}. In the original Bargmann-Wigner equations, the mass element is associated with a fermion quantum. Here, the   equations describe      two quanta, with   the action
of the derivative operating on same-momentum ket bra  (particle anti-particle)  contributions
 \begin{eqnarray} \label{BargmannWignerProcaMomentum}
 (\partial_\mu   |\psi_u(    p,s_\alpha)\rangle) \langle  \psi_v(    p,s_\alpha)| = \frac{1}{2} [ (\partial_\mu    |\psi_u(    p,s_\alpha) \rangle)\langle  \psi_v(    p,s_\alpha)|+ |\psi_u(    p,s_\alpha) \rangle(\langle \psi_v(    p,s_\alpha)|   \overleftarrow{\partial}_\mu) ].\nonumber\\
\end{eqnarray} Application of
  the Dirac operators as in Eqs.  \ref{DiracEq},  \ref{DiracLeft} from the left and right, respectively,
  lead to the conditions
\begin{eqnarray} \label{BargmannWigner}
  F_{{\scriptscriptstyle{BW} }\mu\nu}=  \frac{1}{2}(\partial_\mu j_{{{\scriptscriptstyle BW} }\nu}- \partial_\nu j_{{\scriptscriptstyle{BW} }\mu}) \\
   \frac{1}{2}\partial^\mu   F_{{\scriptscriptstyle{BW} }\mu\nu}= -m^2 j_{{\scriptscriptstyle{BW} }\nu}.
\end{eqnarray}
  Substitution of the first into the second leads to Proca's equations
  \begin{eqnarray} \label{BargmannWignerProca}
 \partial_\mu  \partial^\mu   j_{{\scriptscriptstyle{BW} }\nu}-  \partial_\nu  \partial^\mu  j_{{\scriptscriptstyle{BW} }\mu}  = -4 m^2 j_{{\scriptscriptstyle{BW} }\nu},
\end{eqnarray}
which reduce to
$ \Box  j_{{\scriptscriptstyle{BW} }\nu}   = -4 m^2 j_{{\scriptscriptstyle{BW} }\nu}$,  as  $\partial^\mu  j_{{\scriptscriptstyle{BW} }\mu}=0$.

The  association is to a fermion and anti-fermion connecting through the amplitudes
\begin{eqnarray} \label{TraceFund}
 F_{{\scriptscriptstyle{BW} }}^{\mu\nu}={\rm tr}   \Psi_{{\scriptscriptstyle{BW} }}  {\sigma^{\mu\nu}}=
   {\rm tr}      \psi     \psi^t   C{\sigma^{\mu\nu}}= \langle \psi_v(    p,s_\alpha )|\gamma_{0 }   \sigma ^{\nu\mu}|\psi_u (    p,s_\alpha ) \rangle\\
  j_{{\scriptscriptstyle{BW} }}^\mu={\rm tr}  \Psi_{{\scriptscriptstyle BW} }  \gamma^\dagger_{\mu} =    {\rm tr}   \psi     \psi^t C \gamma^{\mu } = \langle \psi_v(    p,s_\alpha)|\gamma_{0 }\gamma^{\mu }   |\psi_u(    p,s_\alpha) \rangle ,
\end{eqnarray}
 using    the trace    and the charge-conjugation properties:
 $\langle \psi_v(    p,s_\alpha)|\gamma_0=  (C  |\psi_u(     p,s_\alpha)\rangle )^t $, and the $\gamma$-matrices order is chosen so as to fit the state matrix elements.

\renewcommand{\theequation}{D\arabic{equation}}
 \setcounter{equation}{0}
\subsection*{Appendix  D: Massive vector}
We normalize   solutions $A^\psi_\mu$ to the derived equation from Eqs. \ref{ApsiJpsi} and \ref{KleinGordon}
\begin{equation} \label{KleinGordonApsi}
  [{\square} + (2 m)^2]A^\psi_\mu=0.
\end{equation} The classical counterpart is
$A^\psi_{ \mu cl}=    \frac{1}{\sqrt{2}}  (A^\psi_{ \mu}+{A^\psi_{ \mu}}^*)$,
and each component  corresponds to a quantum's creation/annihilation.

 Eq. \ref{DiracCurrenttildeTrans} presents a  massive on-shell vector $A^\psi$,  with  momentum  $2 p_z$, and circular polarization along the transverse directions $\hat {\bf x}$, $\hat {\bf y}$  is  assumed  enclosed
 in a volume $V$  \begin{equation}
  \label{vector}
 A^\psi_\mu =     N^\psi   \varepsilon_\mu  e^{-i 2  k x }, \ \  \\
 \end{equation}
where   the momentum is $2p_z=2\hbar k_z$,   $2kx=2(Et-p_z z)/\hbar$, $\varepsilon_\mu=\frac{1}{\sqrt{2}}(0,1,i,0)$,  $\partial^\mu A^\psi_\mu=0$ as required,
  and $N^\psi= \sqrt{\frac{\hbar c}{  2 E V }}$ is fixed from the normalization condition in   Eq. \ref{normalization}.

  The electric and magnetic fields are, using Eq. \ref{ElectricMagnetic},
  \begin{eqnarray}
  {\bf E} ^\psi  & = & ( F^{\psi 01}, F^{\psi 02}, F^{\psi 03})   =  -(\partial _t  A^{\psi1}+ \partial_x A^{\psi0} , \partial _t   A^{\psi2}+ \partial _y A^{\psi0}  ,\partial _t  A^{\psi3}+ \partial _z A^{\psi0} )  \nonumber
   \\
  {\bf B}^\psi   & = & ( F^{\psi 32},-F^{\psi 31}, F^{\psi 21})  =   (\partial _y  A^{\psi3}- \partial _z  A^{\psi2} , \partial _z  A^{\psi1} -\partial _x   A^{\psi3} ,\partial _x  A^{\psi2}- \partial _y  A^{\psi1} ).  \nonumber  \\ \label{EleMag}
\end{eqnarray}

This  gives for  Eq.  \ref{vector},
  \begin{eqnarray}
  {\bf E}  ^\psi
      & = & \frac{2 N ^\psi  E }{  c \hbar   }
 e^{-i 2 k x}\frac{1}{\sqrt{2}} (i ,-1,0)   \\
  {\bf B} ^\psi
   & = &  \frac{2  N ^\psi   p_z}{    \hbar  }
e^{-i 2   k x}\frac{1}{\sqrt{2}} (1 ,i,0).
\end{eqnarray}


$A^s _\mu$ is similarly obtained, while its mass term is constructed using \ref{ApsiJpsi} and \ref{AsScales}, leading to the same equation as  \ref{KleinGordonApsi},
with solution  \begin{equation}
\label{vectorAs}
 A^\psi_\mu =     N^s   \varepsilon_\mu  e^{-i 2  k x }, \ \  \\
 \end{equation}
This expression
 contains  $\hbar$, while in principle, one can revert to  classical variables using the established equivalences for  the frequency $\omega$ and  the wave number $k$
 in terms of the energy and momentum,
  ${E} = \hbar {\omega}$,  ${\bf p} = \hbar {\bf k}  $. We follow Eq. \ref{energynormalizationMax}, finding $N^s= \hbar c \sqrt{\frac{1}{2 E V }}$.

\renewcommand{\theequation}{E\arabic{equation}}
 \setcounter{equation}{0}

  \subsection*{Appendix  E:
  Electromagnetic unit independence of charge definition}

The obtained charge in Eq. \ref{JaimeConstant} is a unit-independent result. Instead of the $g $ factor $4  \pi$ for  $j_\mu$   in   Maxwell's Eq. \ref{Maxwell},      charge units  $g_f= f g$     require   $  A^s_{f\mu}=(1/f) A^s_\mu$ to maintain the Lorentz force,  so Eq. \ref{Maxwell} is rewritten
\begin{equation} \label{MaxwellGauss}
 {\partial}^\nu  ( {\partial}_\nu A^s_{f\mu}      -{\partial} _\mu  A^s_{f\nu}  )
  =  \frac{4 \pi}{f^2 }  g_f\bar\psi  \gamma_\mu \psi  .   \end{equation}
  The combined expression giving the energies in Eqs.
   \ref{energynormalizationMax} is unmodified,     $ \frac{4 \pi}{f^2 } g f    A^\psi _{\mu}=  \frac{4 \pi}{f^2 }  g_{f }  A^\psi =   -\frac{1}{f  }A^s_{\mu} =- A^s_{f\mu} $, so $ \frac{4 \pi}{f^2 } g_{f }  A^\psi_{\mu}  = - A^s_{f\mu}$.     Eq. \ref{energynormalizationMax} is converted to
   \begin{eqnarray}
   \label{energynormalizationMaxMod}
 & \int d^3x\frac{f^2}{4\pi} [ \frac{1}{2}(  |{\bf E}_{f}^s  |^2 +   |{\bf B}_{ f}^s |^2) +  \frac{(2 m)^2}{2}   ( 2 {A^s_{ f}}_0   {A^s_{ f}}_0  - g_{00}    {A^s_{ f}} _\mu   {A^s_{  f}}^\mu )]
= 2 E.  
 \end{eqnarray}
    Eq.  \ref {eleDiracVecEquality} transforms to
$ \frac{1}{f  } \sqrt{4 \pi\hbar c} A^\psi _\mu=   -A^s_{f\mu}$, leading to
   $ g_f/\sqrt{\hbar c}=\frac{f}{\sqrt{4 \pi}}$. $f=\sqrt{4 \pi}$ converts Gaussian to Lorentz-Heaviside units. This demonstration can be summarized by tracing the $4 \pi$ in Eqs.  \ref{Maxwell}, \ref{energynormalizationMax}, as the substitution  $4 \pi\rightarrow 4 \pi/f^2$ extends it to arbitrary units.

\renewcommand{\theequation}{F\arabic{equation}}
 \setcounter{equation}{0}
  \subsection*{Appendix  F: \bf Tensor extended identities}

  Dirac's
Hermitian conjugate    equation is
\begin{equation} \label{DiracEqHerConj}
 \psi^\dagger (x)  \{   [-i\overleftarrow{\partial}_{\mu}-g_e A_{\mu}(x) ]\gamma^{0}\gamma^{\mu}-m\gamma^{0}\} =0.
\end{equation}

To obtain Eq. \ref{Current}, we sum the equations obtained by   partial contraction:   multiplying Eq. \ref{DiracEq}     by $\bar \psi \gamma_\mu$, $\bar  \psi=\psi ^\dagger \gamma_0$,    from the left, and   Eq. \ref{DiracEqHerConj} by $\gamma_\mu\psi$ from the right. Using   {Appendix A} (Eq. \ref{2gamiden}),  we get   the  generalized Gordon  identity\cite{Gordon} for the interactive case quoted.

This identity can be extended to the antisymmetric component. The third term  contains  another  superconductivity component.
To rewrite this term with single $\gamma^\mu$ matrices, one relies  on the identity derived
by multiplying
 Eq. \ref{DiracEq} by $\bar \psi \gamma_\mu \gamma_\nu$ and Eq. \ref{DiracEqHerConj} by $  \gamma_\nu \gamma_\mu \psi$,     and subtracting the expressions. One finds
\begin{equation}\label{GeneGordonTensor}
 \epsilon_{\mu\nu\eta\sigma}\bar \psi i\overleftrightarrow{\partial}^\eta\gamma_5 \gamma^\sigma \psi +\epsilon_{\mu\nu\eta\sigma} 2\bar \psi\gamma_5 \gamma^\sigma  \psi g_e A^{\eta}+
 {\partial}_\mu  \bar\psi  \gamma_\nu \psi-{\partial}_\nu  \bar\psi  \gamma_\mu \psi
  = m i\bar\psi   [ \gamma_\mu,\gamma_\nu] \psi,
\end{equation}
where we apply Eq. \ref{3gamiden},  and $\epsilon_{\mu\nu\eta\sigma}$ is the Levi-Civita tensor, given in  Eq. \ref{LeviCivita}.
The combination of this equation and Eq. \ref{Current} shows  the Bargmann-Wigner in Eq. \ref{semiMaxwell} as a particular case with a Maxwell's equations structure.

\end{document}